\documentclass[twocolumn,pre,showpacs]{revtex4}

\usepackage{epsfig}

\begin{document}

\title{Instability of defensive alliances in the predator-prey model on complex networks}

\author{Beom Jun Kim}
\email{beomjun@ajou.ac.kr}
\altaffiliation{Present address: Department of Physics, Sung Kyun Kwan University, Suwon 440-746, Korea}
\author{Jianbin Liu}
\affiliation{Department of Molecular Science and Technology, Ajou University,
Suwon 442-749, Korea}
\author{Jaegon Um}
\author{Sung-Ik Lee}
\affiliation{National Creative Research Initiative Center for Superconductivity, 
POSTECH, Pohang 790-784, Korea}

\begin{abstract}
A model of six-species food web is studied in the viewpoint of
spatial interaction structures. Each species has two predators and
two preys, and it was previously known that the defensive
alliances of three cyclically predating species self-organize 
in two-dimensions. The alliance-breaking transition occurs 
as either the mutation rate is increased or interaction topology
is randomized in the scheme of the Watts-Strogatz model.
In the former case of temporal disorder, via the finite-size scaling analysis
the transition is clearly shown to belong to the two-dimensional  
Ising universality class. In contrast, the geometric or spatial 
randomness for the latter case yields a discontinuous phase transition.
The mean-field limit of the model is analytically
solved and then compared with numerical results. The dynamic
universality and the temporally periodic behaviors are also discussed.
\end{abstract}
%89.75.Hc Networks and genealogical trees
%89.75.Fb Structures and organization in complex systems
%64.60.Ht Dynamic critical phenomena
%87.23.Cc Population dynamics and ecological pattern formation

\maketitle

The rock-scissors-paper (RSP) game~\cite{ecoli} gives a typical
three-strategy model of cyclical predator-prey food chain if 
rock, scissors, and paper are replaced by different competitive
species in biology. In this simple game, a rock beats a pair of
scissors, a pair of scissors beat a sheet of paper, and a sheet of
paper beats a rock. As generalizations of RSP game, the food webs
composed of different number of species show a variety of interesting
behaviors~\cite{pre54-6186}. Especially, the models 
proposed by Szab\'o \textit{et al}. in Refs.~\cite{szabo1} 
and~\cite{szabo2} have been shown to self-organize to
form the defensive alliances, which becomes  unstable as the mutation rate
is increased. 

\begin{figure}
\includegraphics[width=0.15\textwidth]{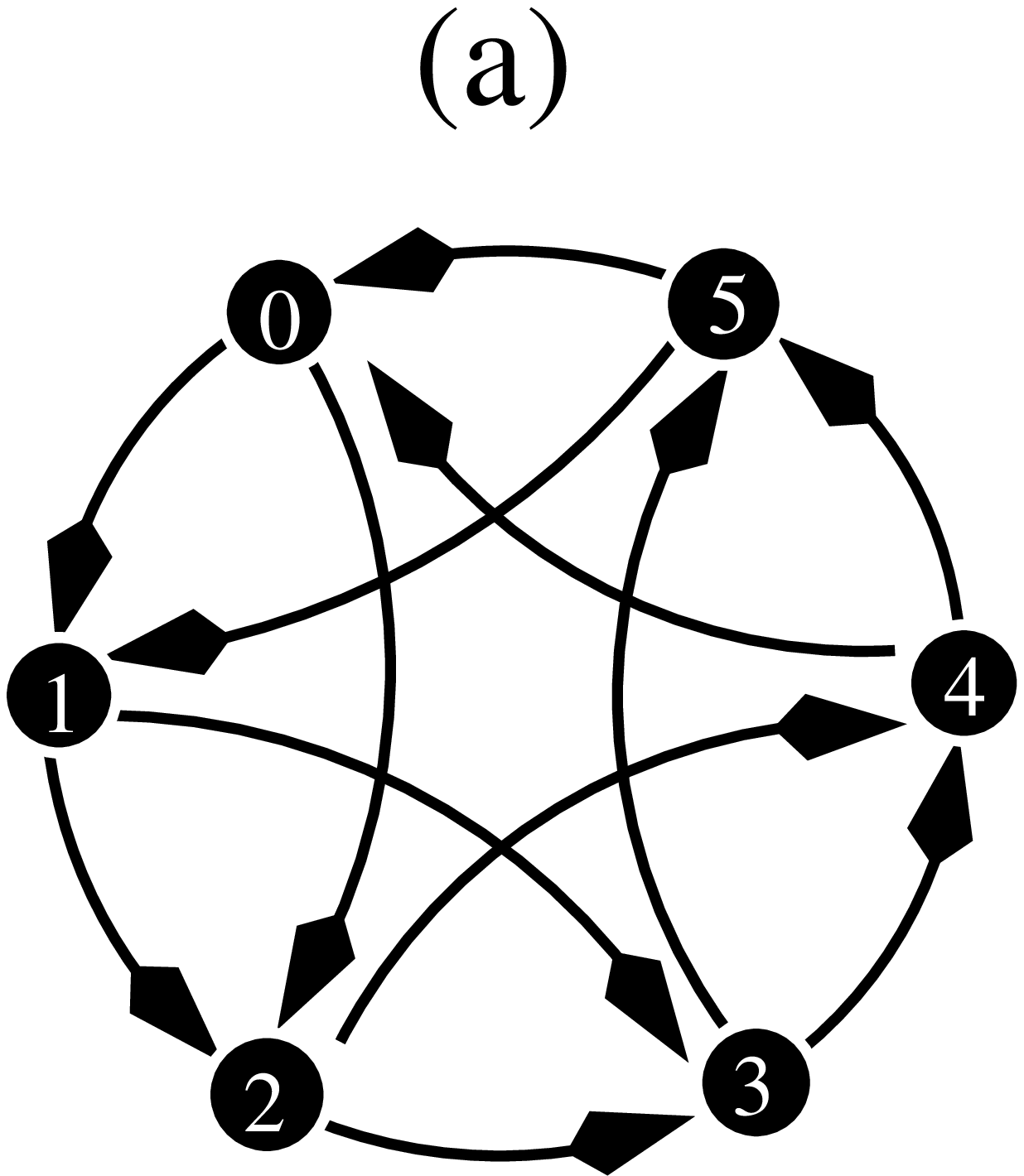} \includegraphics[width=0.15\textwidth]{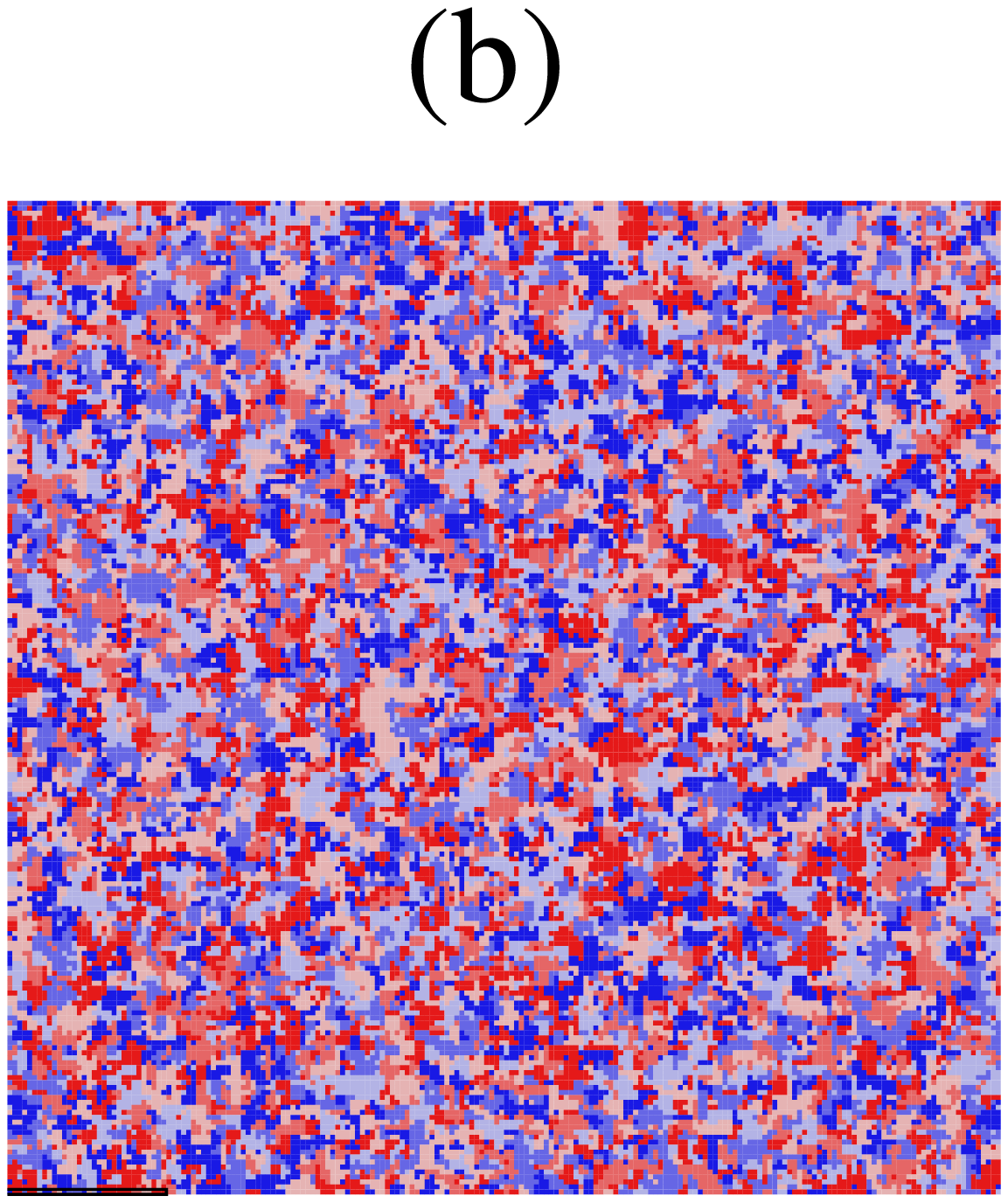}\includegraphics[width=0.15\textwidth]{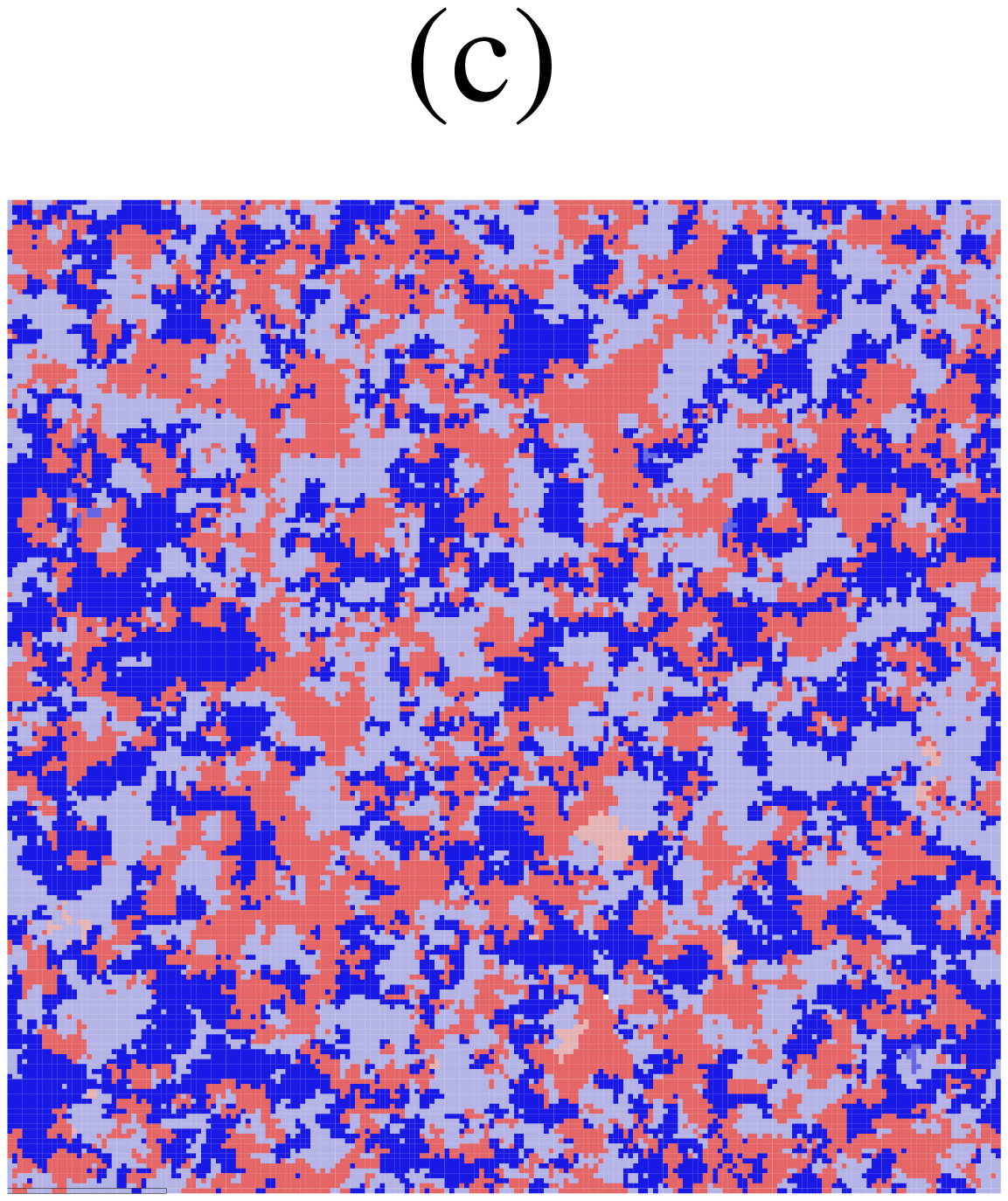}
\caption{(a) The interaction topology used in this work
(Model A in  Ref.~\cite{szabo1}). Arrow e.g., from 0 to 1
indicates that the species 0 is a predator to the species 1
(and thus the species 1 is the prey to the species 0). 
As time goes on the initial random configuration in (b)
evolves to (c), where only three species [(0,2,4) or (1,3,5)] exist
to form the defensive alliance. Different colors indicate
different species in (b) and (c). (Color online.) } \label{figa}
\end{figure}

In the present study, we use the model A in Ref.~\cite{szabo1}
described by the interaction topology in Fig.~\ref{figa}(a),
where the six species weave a food web that is more complicated than
the RSP game: Each species has two predators, two
preys and one neutral interacting partner. 
There indeed exists ecological systems showing the cyclic dominance:
Examples include the three-morph mating system in the side-blotched 
lizard in which each morph dominates another morph when rare~\cite{lizard}. 
Another example is the system of three different populations
of {\it Escherichia coli}~\cite{ecoli} composed
of  toxin-producing (T), toxin-resistant (R), and toxin-sensitive (S) 
strains.  If the growth rate of each group satisfies
S $>$ R $>$ T, the RSP game captures the cyclic dominance in
the system: T dominates S (S is killed by T), S dominates R due to the
higher growth rate, and R dominates T from the same reason.
If bacteria produce two different toxins, the food web is constructed
from nine different species as has been well studied in Ref.~\cite{szabo2}.

The main interest here
is to study the effect of the spatial interaction structure
and for that purpose we play the predator-prey game on the complex network 
structure of the Watts-Strogatz (WS) network~\cite{WS} 
constructed on two-dimensions (2D) as follows:
(1) We first build the 2D $L\times L$ ($N \equiv L^2)$ regular square lattice.
(2) Every bond is visited once and with the rewiring probability $\alpha$
is rewired to the randomly chosen other site.
The above procedure then yields a network structure which possesses
characteristics such as the short characteristic 
path length~\cite{WS,networkreview}.

Once the network is built as described above, the time evolution  of the system 
obeys the following rules~\cite{szabo1}: (1) Six
species are scattered randomly on a square lattice as 
in Fig.~\ref{figa}(b), then (2) the species on each randomly chosen 
site is mutated to one of its predating species 
with the mutation rate $P$. 
(3) If no mutation occurs (with the probability $1-P$), one of the 
nearest neighbors is chosen to interact, and the dominant one
survives and invades the subordinated one. For example,
if the pair of species 0 and 1 are chosen, the species 1 is 
replaced by the species 0 [see Fig.~\ref{figa}(a)].  
If two neutral partners has been chosen, i.e., no arrow connects
the two species in Fig.~\ref{figa}(a), nothing happens.
For the 2D regular square lattice corresponding
to the rewiring probability $\alpha=0$, it has been found that
the defensive alliances composed of three species, 
(0,2,4) or (1,3,5), are spontaneously formed at a small mutation rate
and the other species that does not belong to the alliance dies 
out~\cite{szabo1} [see Fig.~\ref{figa}(c)].
As the mutation rate is increased, it has been shown~\cite{szabo1}
that the defensive alliance
becomes unstable and there occurs well-defined phase transition
of the universality class of the 2D Ising model~\cite{textbook},
which is expected because of the existence of the $Z_2$ symmetry:
Interchange of two alliances (0,2,4) $\leftrightarrow$ (1,3,5)
do not change the game rules, which we call here {\it inter-alliance
symmetry}.
The spontaneous breaking of the defensive alliance originates
either from the high mutation rate $P$ or from the high degree 
of structural randomness controlled by the rewiring probability $\alpha$.
In other words, the instability of the defensive alliance
is induced either by the temporal randomness or by the structural
randomness. In reality, these two different types of randomness
may coexist. However, from the practical computational difficulty
we in this paper only investigate the effect of each randomness separately.

We first investigate the phase transition for the 2D regular 
square lattice corresponding to the WS network with the rewiring 
probability $\alpha=0$.
Although this was previously studied~\cite{szabo1},
we in this work use the extensive finite-size scaling analysis to 
confirm not only the 2D Ising static universality class but also
to identify the dynamic universality class detected
by the dynamic critical exponent $z$.
In order to describe the alliance breaking transition
we define the order parameter which functions as a 
magnetization in the Ising model:
\begin{equation} \label{eq:m}
m=(c_0+c_2+c_4)-(c_1+c_3+c_5),
\end{equation}
where $c_s \equiv N_s/N$ $(s=0,\cdots ,5)$ is the 
density of species $s$ with $N_s$ being
the number of sites occupied by $s$.
When $P$ is sufficiently small, the defensive alliance is
formed with $c_0 \approx  c_2 \approx  c_4 \approx  1/3$ and $c_1 \approx  c_3 \approx  c_5\approx   0$ 
or vice versa, leading to $m \approx \pm 1$ (ordered phase). 
When the system recovers its full symmetry at a high mutation rate, 
all species have the same density,  yielding $m \approx 0$.
Only for convenience we use 
\begin{equation}
\mu \equiv \ln(1/P)
\end{equation}
as a control parameter. High mutation rates correspond to the
small values of the mutation parameter $\mu$, and thus qualitatively speaking,
the physical meaning of $\mu$ resembles that of the 
inverse temperature in the standard statistical mechanics.

Numerical simulations are performed on systems of sizes 
varying from $L=32$ to 192 with  the periodic boundary conditions 
employed. After equilibration for $10^4$ steps per site, which
is sufficiently long enough, the thermal average is computed 
for later $10^4$ steps at least. 
The measured quantities are $\langle |m| \rangle$,
the Binder's cumulant~\cite{binder}
\begin{equation}
U_L \equiv 1-\frac{\langle m^4 \rangle}{3\langle m^2 \rangle^2},
\end{equation}
and the susceptibility 
\begin{equation}
\chi \equiv N \bigl(\langle m^2\rangle-\langle |m|\rangle^2 \bigr),
\end{equation}
where $\langle...\rangle$ denotes the thermal average. 
In order to study dynamic critical behavior, we also measure
the autocorrelation function as a function of time $t$ defined by
\begin{equation} \label{eq:Ct1}
C(t) \equiv \langle |m(t)m(0)|\rangle -\langle |m|\rangle^2 .
\end{equation}

The finite-size scaling forms of measured quantities are written
as~\cite{FSS}  
\begin{eqnarray}
& & \langle |m| \rangle  = L^{-\beta/\nu}\widetilde{m}\bigl((\mu-\mu_c) L^{1/\nu}\bigr), \label{eq:mscale} \\
& & U_L  =  \widetilde{U}\bigl((\mu-\mu_c)L^{1/\nu}\bigr), \\ 
& & \chi  =  L^{\gamma/\nu}\widetilde{\chi}\bigl((\mu-\mu_c)L^{1/\nu}\bigr), \\
& & C(t)/C(0) =  \widetilde{C}(t L^{-z}),  \label{eq:Ct2}
\end{eqnarray}
where $\widetilde{m},\widetilde{U},\widetilde{\chi}$, and $\widetilde{C}$ 
are suitable scaling functions, $\mu_c$ is the critical value of $\mu$,
and $\beta$, $\gamma$, $\nu$ are standard critical exponents~\cite{textbook}
while $z$ is the dynamic critical exponent defined at $\mu_c$ from
the divergence of the relaxation time scale ($\tau \sim L^z$).

\begin{figure}
\includegraphics[width=0.25\textwidth]{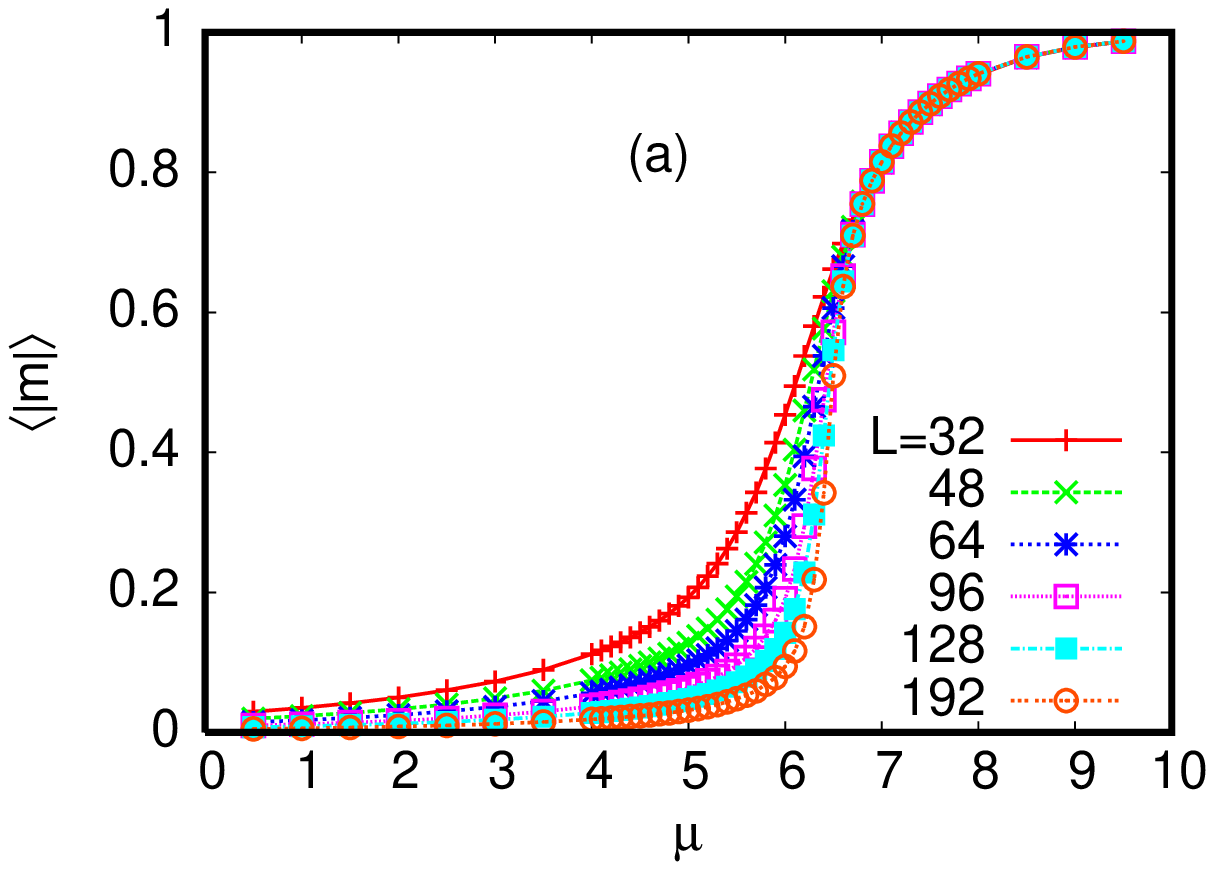}\includegraphics[width=0.25\textwidth]{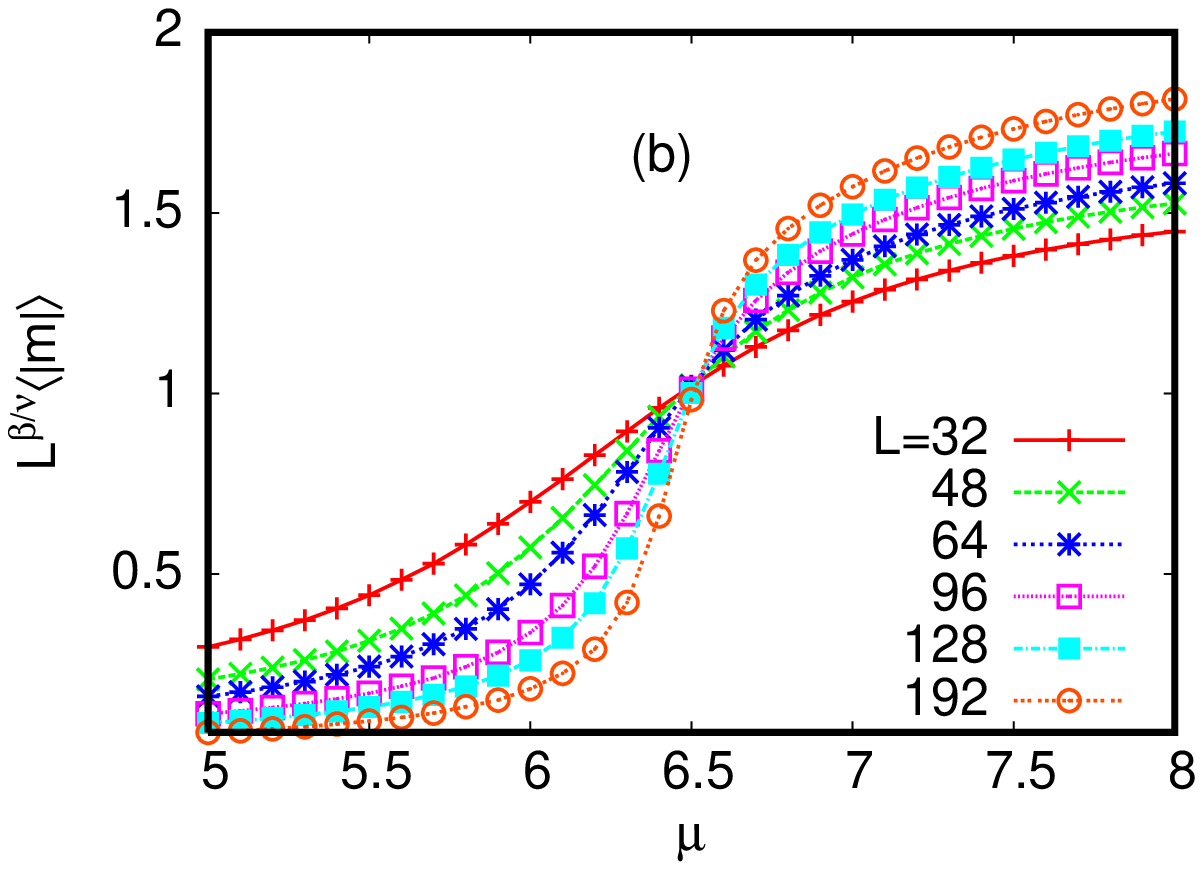}

\includegraphics[width=0.25\textwidth]{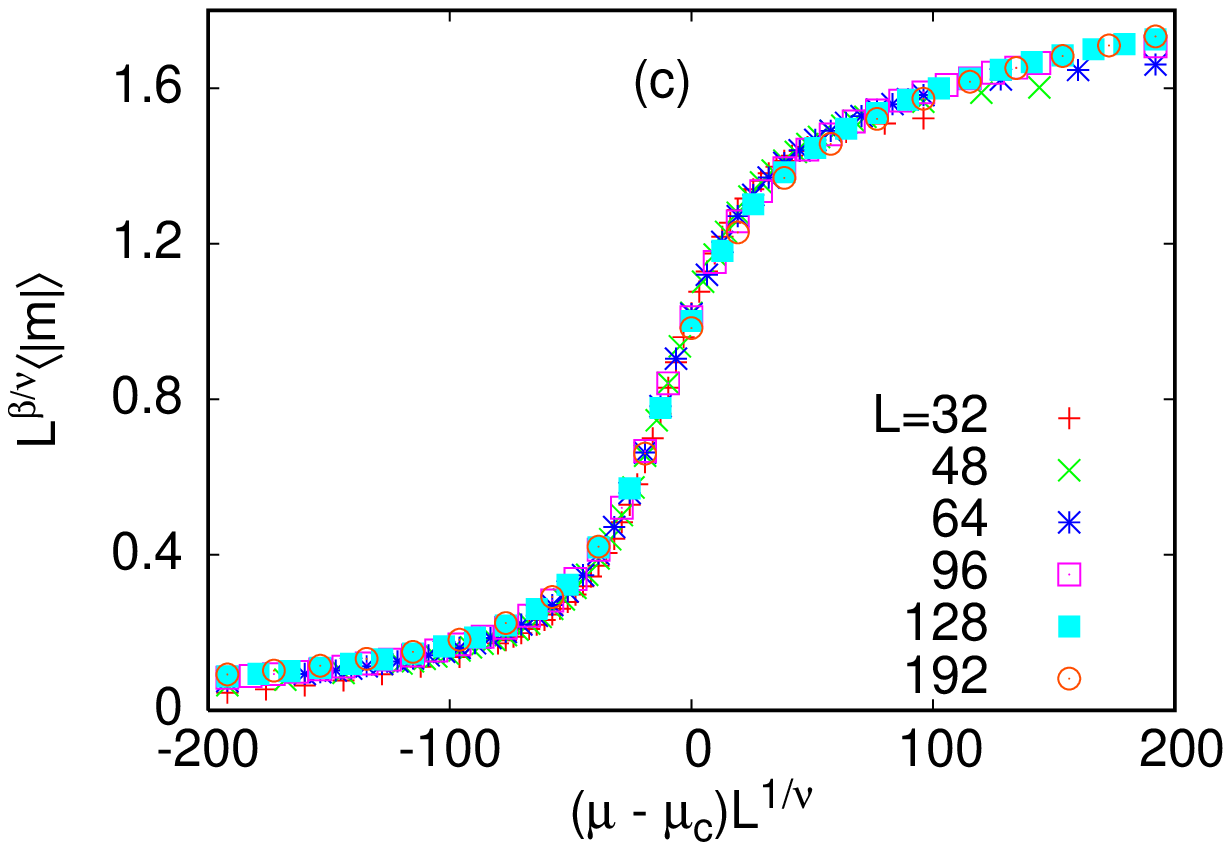}\includegraphics[width=0.25\textwidth]{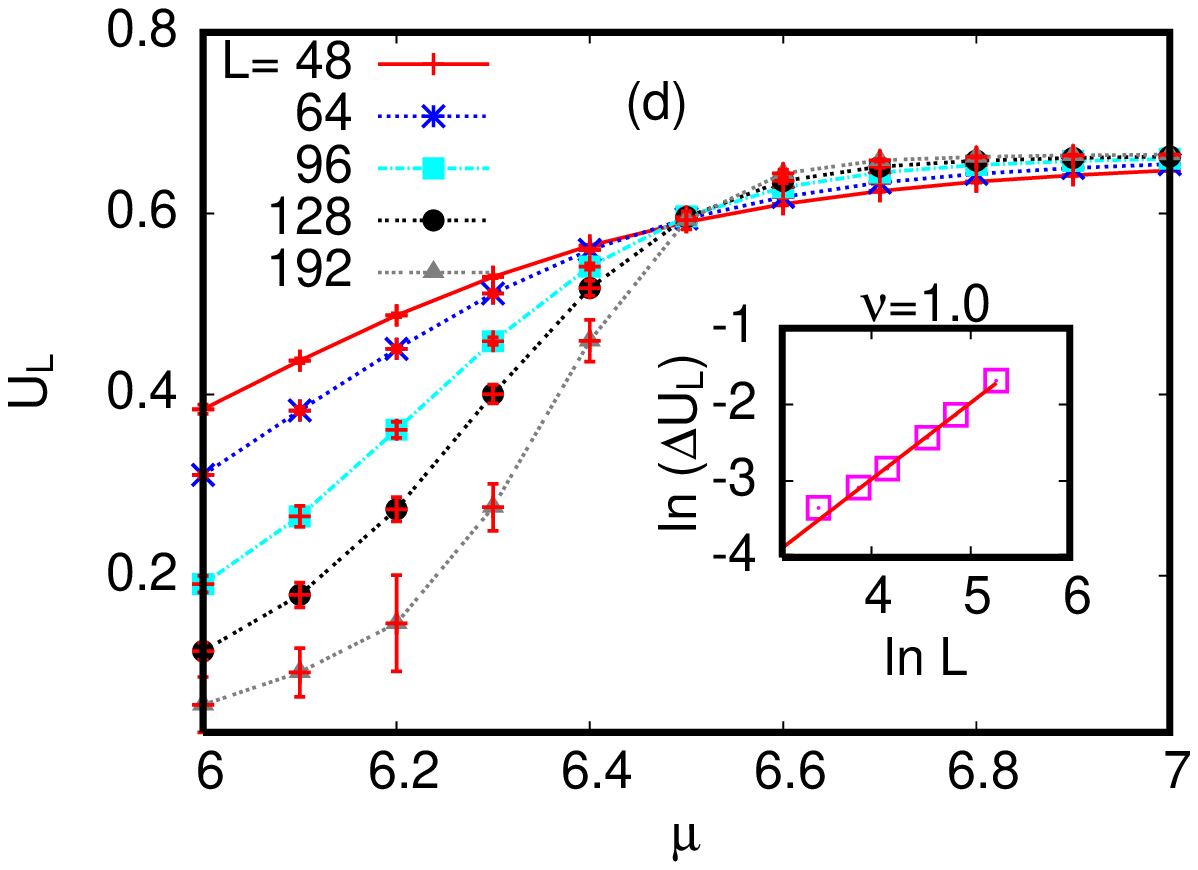}

\includegraphics[width=0.25\textwidth]{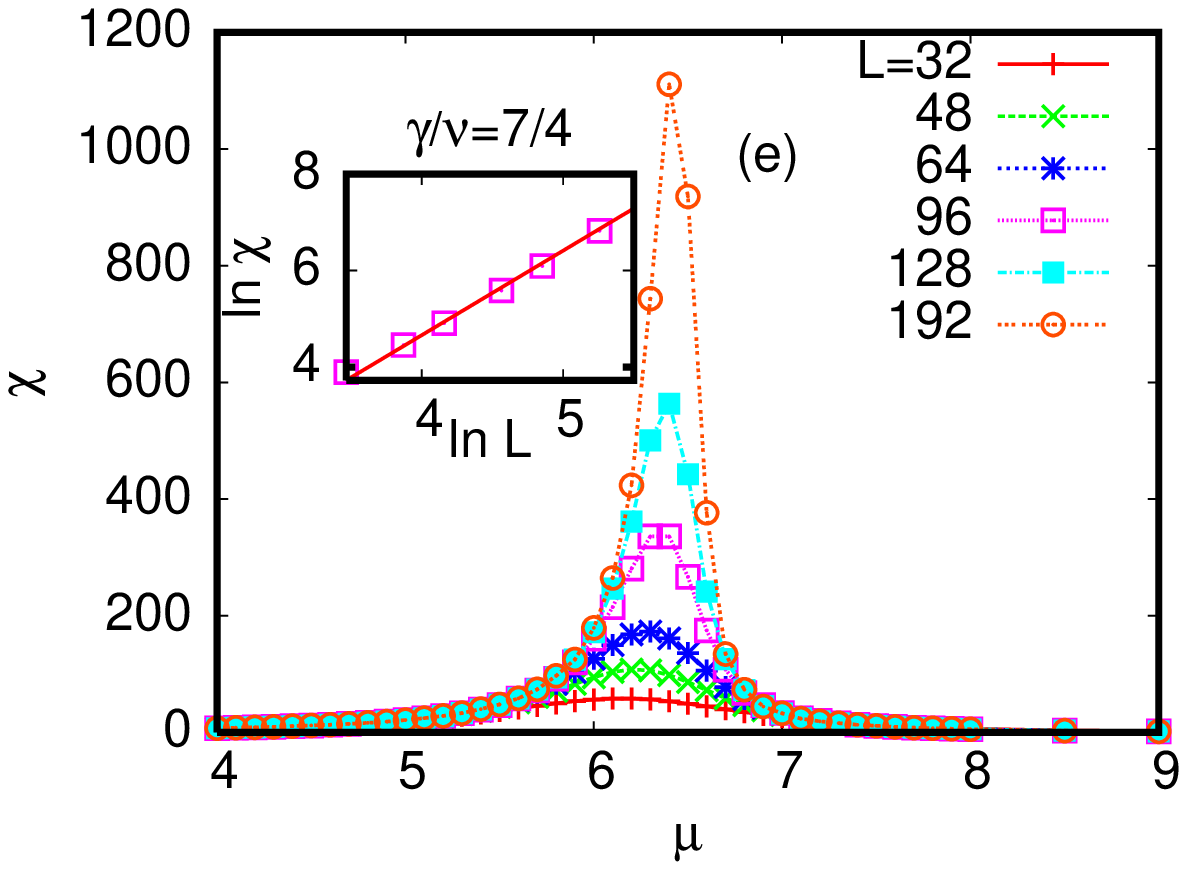}\includegraphics[width=0.25\textwidth]{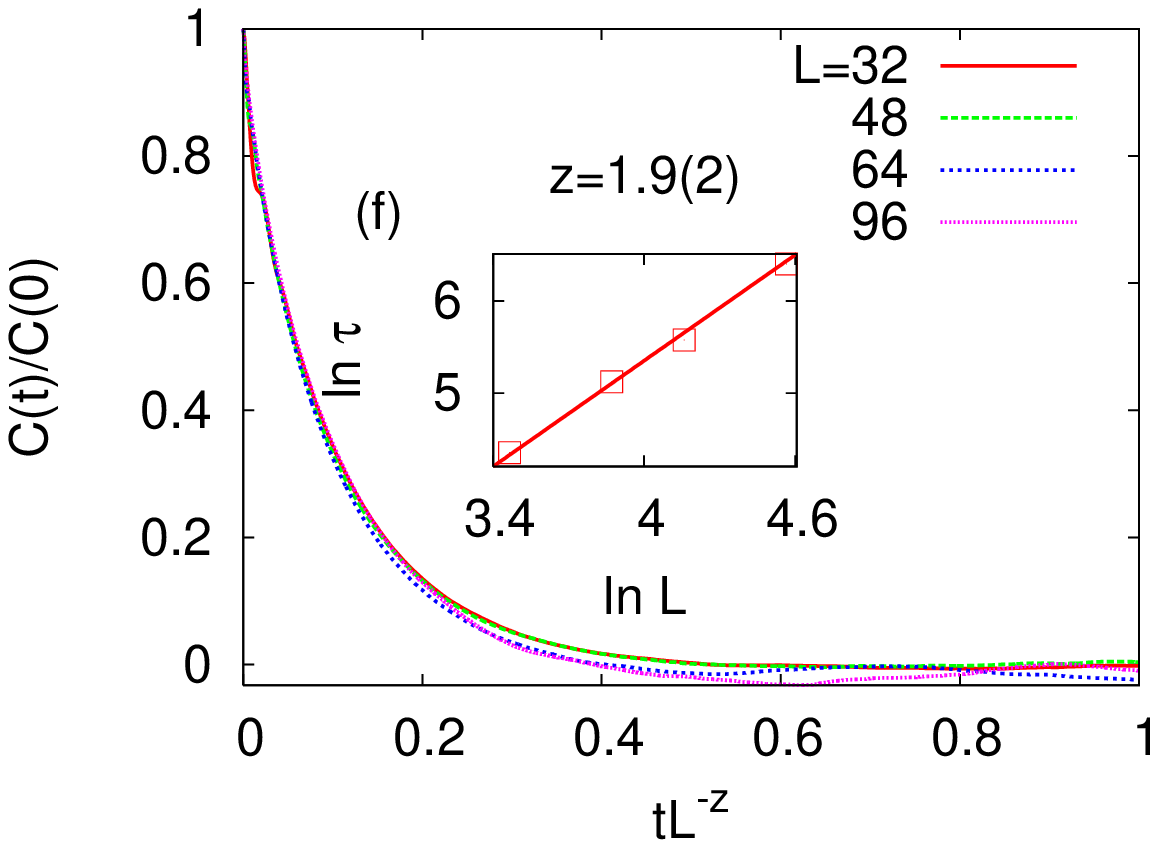}

\caption{ Phase transition in the 2D regular lattice in terms
of the mutation parameter $\mu$. (a) The order parameter $\langle |m| \rangle$
versus $\mu$ clearly shows the existence of the alliance breaking transition.
(b) A unique crossing at $\mu_c=6.50(4)$ with $\beta/\nu\approx1/8$ is obtained
from the finite-size scaling form of $\langle |m| \rangle$. (c) All data points
for $\langle |m| \rangle$ in (b) collapse into to a smooth curve by using
scaled variables. (d) Binder's cumulant at different sizes cross at $\mu_c=6.50(4)$
in agreement with (b). The inset in (d) shows that
$\Delta U_L \sim L^{1/\nu}$ to yield $\nu \approx 1$, where $\Delta U_L
\equiv  U_L(\mu_1) - U_L(\mu_2)$ with $\mu_{1,2}$ are two adjacent
points near $\mu_c$.
(e) The susceptibility $\chi$. The inset shows a
log-log plot of $\chi$ versus $L$ at $\mu_c$ leading to 
$\gamma \approx 7/4$. (f) The autocorrelation function $C(t)/C(0)$ 
at $\mu_c$ versus $tL^{-1.9}$ with $z=1.9$. All curves for
different sizes collapse well to a single curve, indicating
that the dynamic critical exponent $z=1.9(2)$. (Color online)} \label{fig2d}
\end{figure}

Figure~\ref{fig2d} summarizes the numerical results for the
phase transition in the 2D regular lattice. The order parameter
$\langle |m| \rangle$ shown in Fig.~\ref{fig2d}(a) exhibits 
the existence of the transition, which is analyzed in detail
in Fig.~\ref{fig2d}(b) and (c) by using the finite-size
scaling form in Eq.~(\ref{eq:mscale}). The critical
point $\mu_c = 6.50(4)$ as well as critical exponents $\beta \approx 1/8$
and $\nu \approx 1$ are obtained, which are confirmed again from  
the finite-size scaling of the Binder's cumulant shown in 
Fig.~\ref{fig2d}(d). The divergence of the susceptibility in 
Fig.~\ref{fig2d}(e) is analyzed
to get $\gamma \approx 7/4$. At $\mu = \mu_c$, we compute
the autocorrelation function~(\ref{eq:Ct1}) and plot it in
Fig.~\ref{fig2d}(f) in accord with the scaling form in Eq.~(\ref{eq:Ct2}):
All curves at different sizes collapse well to a single curve
with the dynamic critical exponent $z \approx 2$.
All these findings clearly confirm that the alliance breaking 
transition induced by the mutation belongs to the 2D Ising universality
class as was known in Ref.~\cite{szabo1}.

We next study the phase transition induced by the spatial
randomness introduced via the long-range shortcut in the
WS network model. One can motivate the study along
this direction since in real systems, the spatial
interaction topology among species can be much more complicated
than the nearest-neighbor interaction on a regular square
lattice. 
Distinguished from the regular network, the small-world
network~\cite{WS,networkreview}, which captures characteristics
of many real networks very well, has 
a remarkably small average path length similar to the globally
coupled network (or the mean-field case).
The three-strategy RSP game has been studied on some small-world
networks and the periodic flourishes of three strategies were found
to happen during the time evolution~\cite{szabo3}.

\begin{figure}
\includegraphics[width=0.25\textwidth]{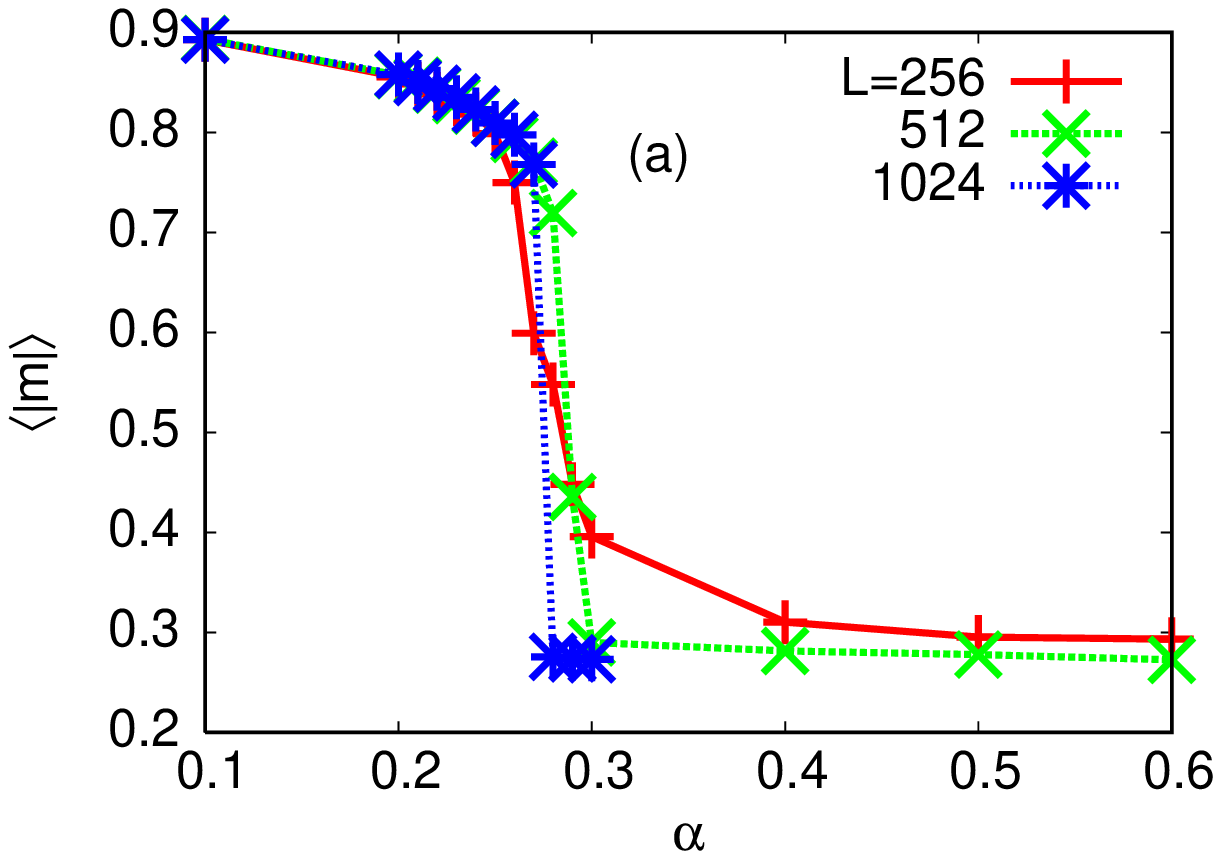}\includegraphics[width=0.25\textwidth]{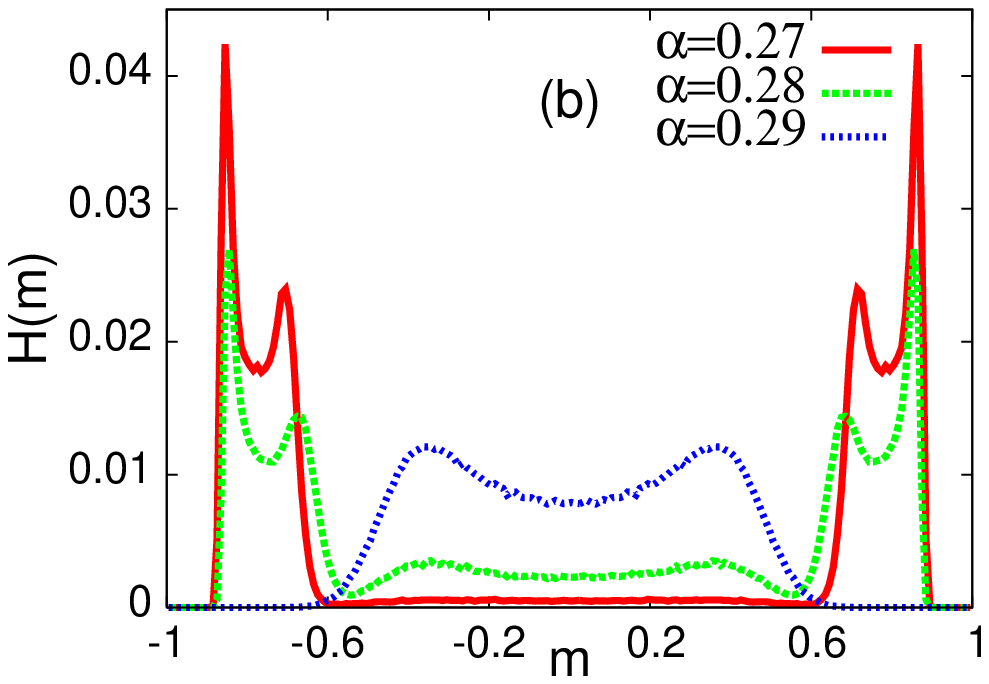}
\caption{Phase transition in the WS network in terms of the rewiring 
probability $\alpha$. The mutation parameter is set to $\mu = 7.0$.
(a) The order parameter $\langle |m| \rangle$ as a function
of $\alpha$ shows a sudden drop at $\alpha_c \approx 0.28$. 
The change of $\langle |m| \rangle$ becomes steeper as $L$ is
increased, indicating that the transition is discontinuous one. 
(b) The normalized histogram $H(m)$ of $m$ for $L=1024$
is displayed at $\alpha = 0.27, 0.28$, and 0.29.
The abrupt change in the form of the histogram between $\alpha = 0.28$ 
and 0.29 clearly confirms again discontinuous transition. (Color online)
} \label{figWS}
\end{figure}

We fix the mutation parameter to $\mu = 7.0$, which is well inside
the ordered phase with the defensive alliance formed
for the 2D regular square lattice (see Fig.~\ref{fig2d}). 
Use of a very large value of $\mu$, corresponding to very small 
mutation rate, was found to make the system approaches a local 
dynamic fixed point and then the system stays there forever.
At $\mu = 7.0$, which is small enough to ensure the equilibration
and large enough to make the system ordered at small $\alpha$,
the system is found to undergo a 
phase transition at $\alpha = \alpha_c \approx 0.28(1)$ 
as displayed in Fig.~\ref{figWS}.
The abrupt drop down of the order parameter $\langle |m| \rangle$
at the transition as displayed in Fig.~\ref{figWS}(a), together
with the change of the histogram $H(m)$, normalized to satisfy
$\sum_m H(m) = 1$,  in Fig.~\ref{figWS}(b), clearly
indicates that the transition is of the discontinuous nature
in a sharp contrast to the finding for 2D regular lattice at $\alpha = 0$
(see Fig.~\ref{fig2d}). In more details, the sudden change of
the peak position of $H(m)$ 
between $\alpha=0.28$ and 0.29 in Fig.~\ref{figWS}(b) 
is interpreted as a strong evidence
of a discontinuous transition. Continuous transition in general
exhibits the continuous shift of the peak position toward $m=0$
as the critical point is approached from the ordered phase.

\begin{figure}
\includegraphics[width=0.25\textwidth]{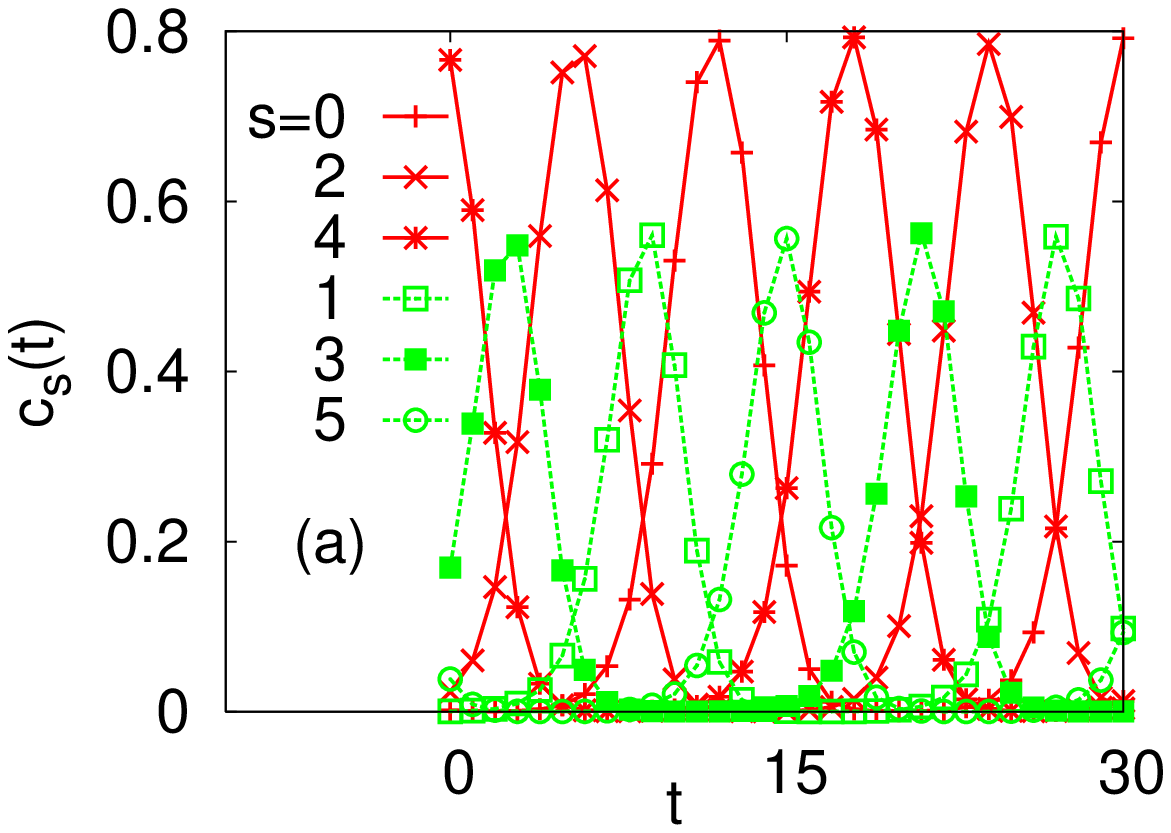}\includegraphics[width=0.25\textwidth]{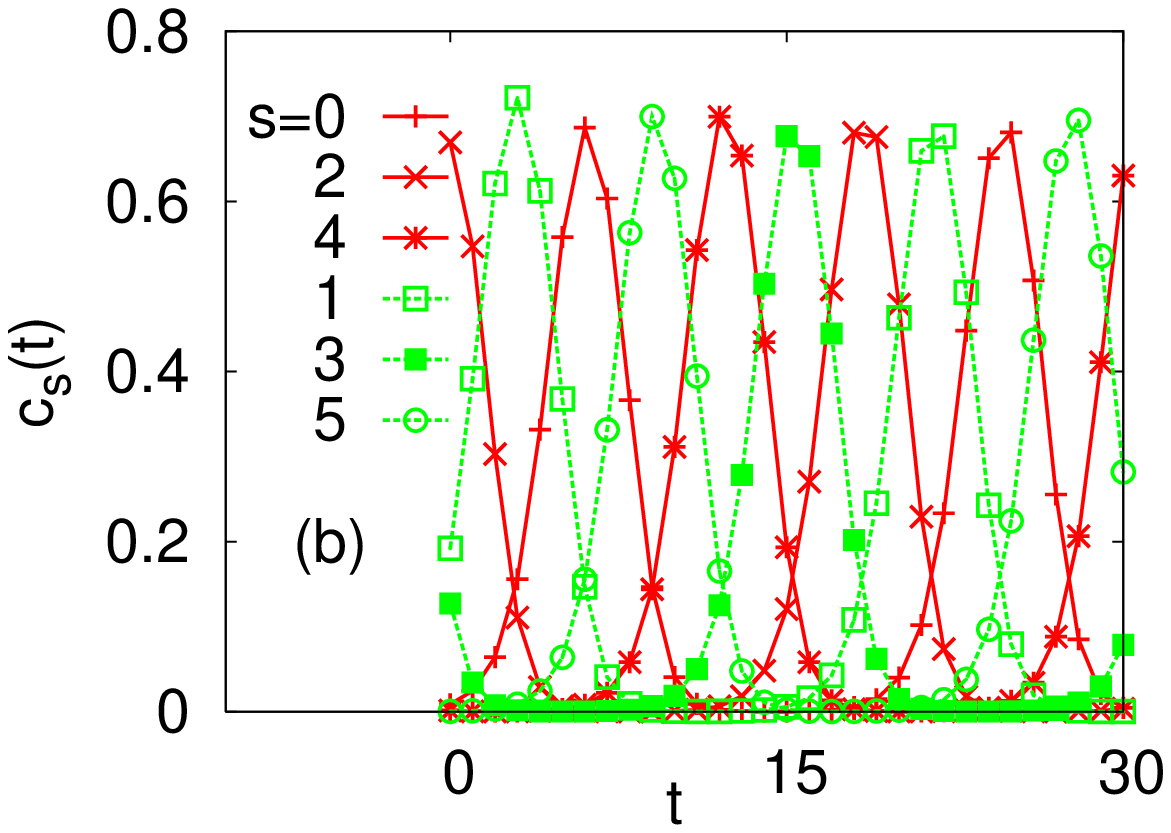}
\caption{Time evolution of density of species $c_s(t)$ for 
the WS network for $L=1024$ at $\mu = 7.0$. (a) At $\alpha = 0.27$, the
inter-alliance symmetry is broken, suggesting $m \neq 0$,
and each species in the alliance dominates all the other
species in a time-periodic fashion. It is interesting to
note that the non-dominant alliance is also formed and
even the member of the non-dominant alliance prevails 
periodically.
(b) At $\alpha = 0.29$, the system recovers its full symmetry
and each species is equivalent to others.
However, the time-evolution of $c_s(t)$
is still periodic, in contrast to the disordered phase 
in 2D regular lattice. The time $t$ is measured
after stationarity is achieved. (Color online)} \label{fig:csWS}
\end{figure}

In general, one can study the phase diagram of the model
in the 2D parameter space of $(\mu, \alpha)$. Due to 
the practical difficulty to cover the whole parameter space,
we in this work only explore the phase transitions
along the two straights line in the $(\mu, \alpha)$ plane:
One on the axis $(\mu, \alpha =0)$, and the other on the
line  ($\mu = 7.0, \alpha)$. Both parameters $\mu$ and $\alpha$
control the amounts of randomness (temporal and spatial ones, respectively),
and accordingly the ordered phase with the defensive 
alliance becomes unstable as either $\mu$ is decreased 
(i.e., the mutation rate $P$ is increased) or $\alpha$ is increased.
Not only the phase transitions have different natures (continuous
one belonging to the 2D Ising universality class for the former, and 
discontinuous one for the latter), but also the ordered
and disordered phases in each case are very much different 
in terms of the time evolution of densities of species. 
In the ordered and disordered phase on the axis of $\alpha=0$, 
the density of each species does not fluctuate much but
stays at almost the same level: For ordered phase at
$\mu > \mu_c$, $c_s \approx 1/3$ for $s \in  \{ \mbox{alliance} \}$
and $c_s \approx 0$ otherwise, while for the disordered phase
at $\mu < \mu_c$, $c_s \approx 1/6$ for all species.
In contrast, the time evolution $c_s(t)$ for the case of the
WS network at $(\mu = 7.0, \alpha)$ is strikingly different.
In Fig.~\ref{fig:csWS}(a), the time evolutions of densities
of species are shown for $\mu = 7.0$ and $\alpha = 0.27 (< \alpha_c)$.
It is clearly shown that the inter-alliance $Z_2$ symmetry
is broken, indicating that the system has a nonzero value
of the order parameter. Another very important observation
one can make from Fig.~\ref{fig:csWS}(a) is that each species
within the alliance ($s=0,2,4$) cyclically dominates 
all the others in a very regular way, and very interestingly,
the species that does not belong to the dominant alliance
also prevails in a time-periodic manner.
Even in the disordered phase at $\alpha > \alpha_c$, 
the periodic dominance persists while the inter-alliance symmetry
is fully recovered [see Fig.~\ref{fig:csWS}(b)].

\begin{figure}
\includegraphics[width=0.25\textwidth]{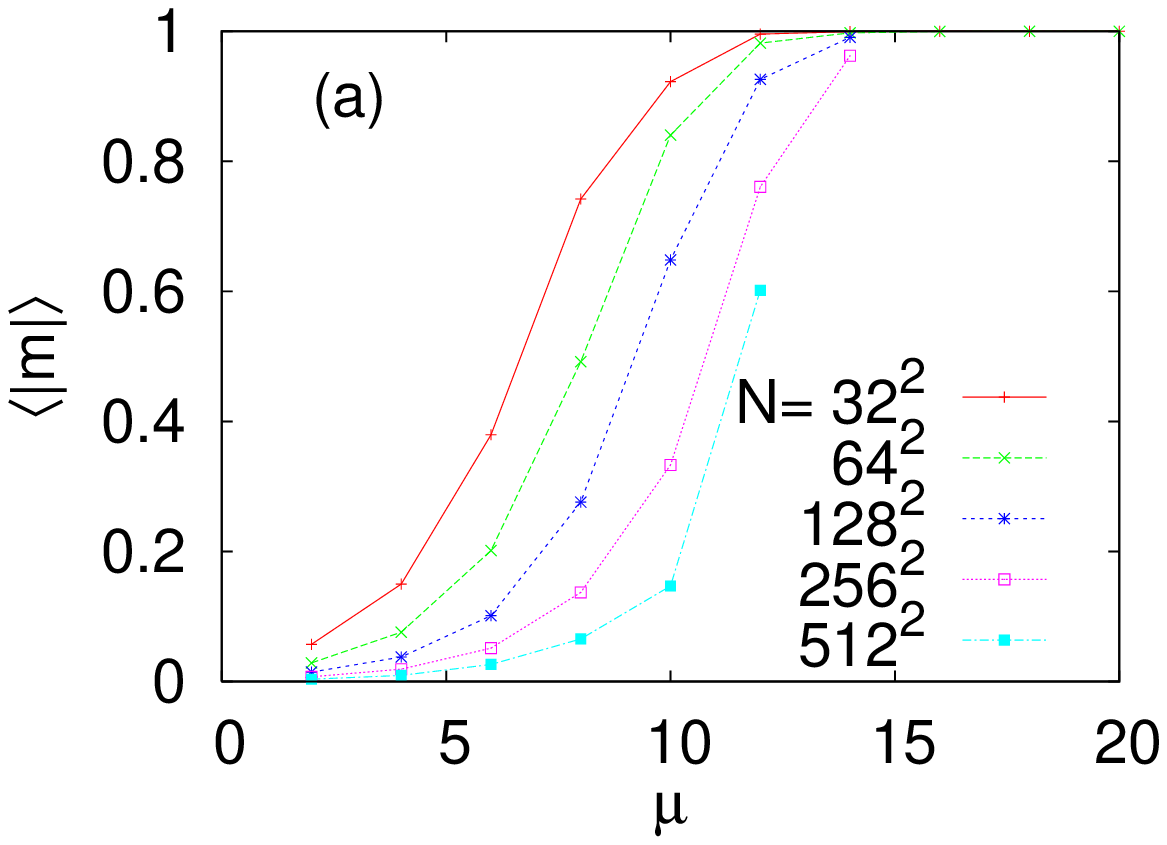}\includegraphics[width=0.25\textwidth]{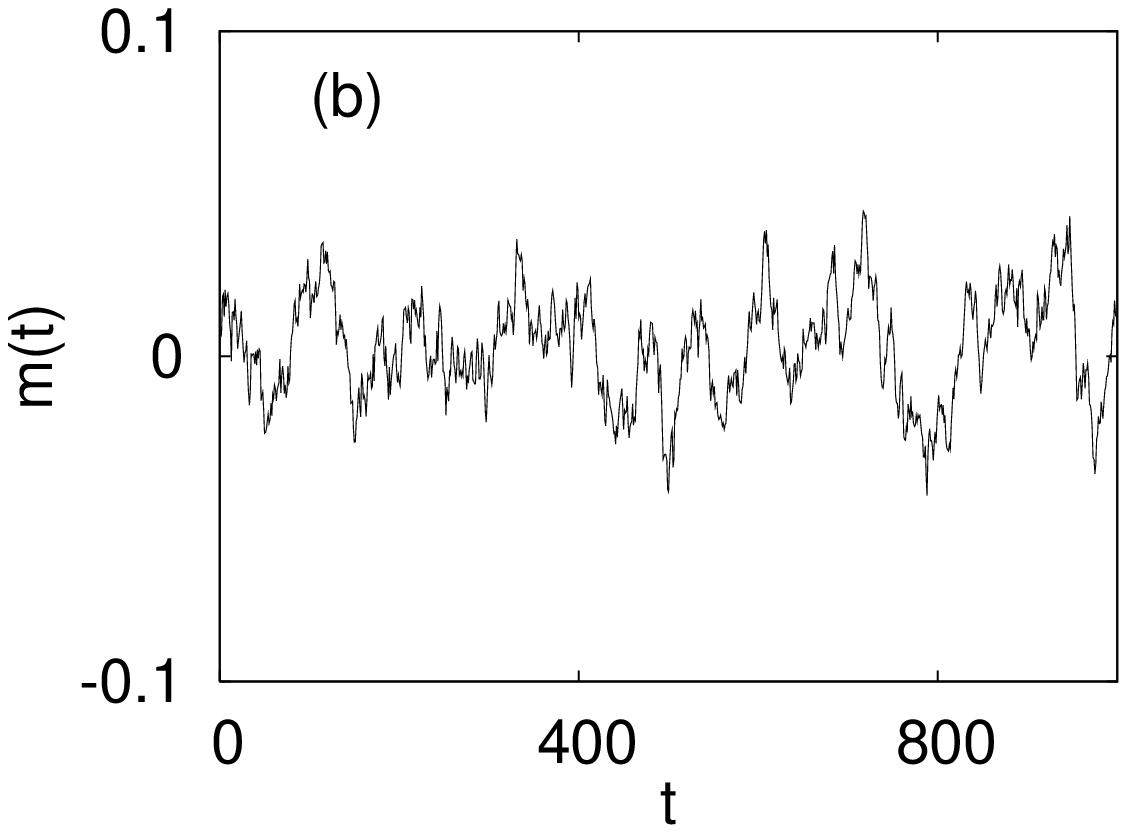}
\caption{Numerical results for the mean-field case.
(a) The order parameter $\langle |m| \rangle$ versus $\mu$ for $N=32^2, 64^2, \cdots, 512^2$. As $N$ is increased the region for the ordered phase
shifts towards higher values of $\mu$, indicating that the
defensive alliance vanishes in the thermodynamic limit at any
nonzero value of the mutation rate. (b) Time evolution of
the order parameter $m(t)$ is not periodic at all, different
from the WS network shown in Fig.~\ref{fig:csWS}. (Color online)}\label{fig:MF}
\end{figure}

We finally investigate the mean-field limit of the 
game~\cite{rspmf}, where all individual species interact with all the
other species in the system. The master equation for the
number $N_s$ of species $s$ is given by
\begin{eqnarray}
\Delta N_s & = & 
P\left(-c_s + \frac{c_{i_1}}{2} + \frac{c_{i_2}}{2}\right) \nonumber \\
&+ &2(1-P)c_s( c_{i_1} + c_{i_2} - c_{j_1} - c_{j_2}) ,
\end{eqnarray}
where $\Delta N_s \equiv N_s(t+1) - N_s(t)$, $i_1$ and $i_2$ ($j_1$
and $j_2$) are two preys (two predators) of $s$. For example,
the species $s=0$ has $i_1 = 1, i_2 = 2, j_1 = 4, j_2 = 5$ [see
Fig.~\ref{figa}(a)].
The  term proportional to $P$ describes the decrease of $N_s$ by 
the mutation from $s$ to its predators and increase of $N_s$ by 
the mutation from its preys to $s$. The other term originates
from the interaction of $s$ to other species: When $s$ meets
its preys (predators) $N_s$ is increased (decreased). The 
numerical factor two in front of $(1-P)$ is due to the interaction
of other species with $s$.
If we start from the situation when $c_0 = c_2 = c_4 = (1 + m)/6$
and $c_1 = c_3 = c_5 = (1-m)/6$, with the order parameter $m$
in Eq.~(\ref{eq:m}), the above master equation is reduced to
a very simple form
\begin{equation}
\frac{dm}{dt} = m(t+1) - m(t) = -NPm, 
\end{equation}
yielding the solution 
\begin{equation}
m(t) = m(0) \exp(-t/\tau)
\end{equation}
with $\tau = PN$.
The mean-field solution indicates that there is no ordered
phase at any nonzero value of the mutation rate, i.e., 
{\em the defensive alliance cannot be formed in the mean-field
limit}. It is also noteworthy that the time-periodic behavior
observed for the WS network in Fig.~\ref{fig:csWS} ceases
to exist in the mean-field case.
The simulation results displayed in Fig.~\ref{fig:MF}
for the mean-field model are in perfect agreements with
the above analytic findings: The ordered phase that
appears to exist for small system sizes drifts away
toward the region of higher value of $\mu$ as $N$ is increased, 
suggesting
the disappearance of the ordered phase in the thermodynamic
limit [see Fig.~\ref{fig:MF}(a)]. There is no time-periodic
behavior of $m$ in equilibrium as shown in Fig.~\ref{fig:MF}(b),
where $t$ is measured after equilibration.

In summery, we have investigated the instability of the
defensive alliances for the simple food web of six
species in three different spatial interaction structures:
the 2D local regular square lattice, the WS network,
and the globally coupled mean-field network.
The temporal randomness imposed by the mutations 
as well as the spatial randomness in the interaction structure
tuned by the rewiring probability in the WS network
has been shown to make the defensive alliance unstable.
When the mutation rate is increased for the 2D square lattice
the alliance breaking transition has been clearly identified
to belong to the 2D Ising universality class due to the
common $Z_2$ symmetry. On the other hand, when the rewiring
probability is increased, the transition becomes discontinuous,
and around the transition the natures of the ordered and disordered
phases are very different from the 2D square lattice.
The mean-field model has also been studied analytically and
numerically with the results that the defensive alliance
cannot be formed at any value of the mutation rate and that
the time-periodic behavior observed in the WS network
is not seen any longer.

B.J.K. was supported by grant  No. R01-2005-000-10199-0
from the Basic Research Program of the Korea Science 
and Engineering Foundation.  J.U. and S.-I.L. 
acknowledge the support by the Ministry of Science and Technology
of Korea through the Creative Research Initiative Program.
The numerical calculations have
been performed in the computer cluster Iceberg at Ajou University.

\end{document}